\documentclass[aps,prl,preprint,groupedaddress]{revtex4-1}
\usepackage{graphicx}
\usepackage{epstopdf}
\usepackage{amsmath}

\usepackage{siunitx}
\usepackage[english]{babel}
\usepackage{dcolumn}
\usepackage{bm}

\begin{document}
\hyphenation{me-tro-lo-gy}
\hyphenation{geo-de-sy}

\title{High accuracy coherent optical frequency transfer over a doubled \SI{642}{km} fiber link}

\author{D. Calonico$^{1}$, E. K. Bertacco$^{1}$, C. E. Calosso$^{1}$,  C. Clivati$^{1}$, G. A. Costanzo$^{1,2}$,  M. Frittelli$^{1}$, A. Godone$^{1}$,  A. Mura$^{1}$, N. Poli$^{3}$, D. V. Sutyrin$^{3}$, G. Tino$^{3}$, M. E. Zucco$^{1}$, F. Levi$^{1}$
}                     
%
%
\address{
$^{1}$Istituto Nazionale di Ricerca Metrologica INRIM, strada delle Cacce 91, 10135, Torino, Italy \\
$^{2}$Politecnico di Torino, Corso Duca degli Abruzzi 24, 10129, Torino, Italy \\ 
$^{3}$Universit\`{a} di Firenze, LENS and INFN, via Sansone 1, Sesto Fiorentino (FI), Italy\\
*Corresponding author: c.clivati@inrim.it
}
\begin{abstract}To significantly improve the frequency references used in radio-astronomy and precision measurements in atomic physics, we provide frequency dissemination through a \SI{642}{km}  coherent optical fiber link, that will be also part of a forthcoming European network of optical links. On the frequency transfer, we obtained a frequency instability of  \SI{3e-19}{} at  \SI{1000}{s}  in terms of Allan deviation on a \SI{5}{mHz} measurement bandwidth, and an accuracy  of  \SI{5e-19}{}. The ultimate link performance has been evaluated by doubling the link to \SI{1284}{km}, demonstrating a new characterization technique based on the double round-trip on a single fiber. The arming of a second fiber is avoided: this is beneficial to long hauls realizations in view of a continental fiber network for frequency and time metrology. The observed noise power spectrum is seldom found in the literature; hence, the expression of the Allan deviation is theoretically derived and the results confirm the expectations. 
\end{abstract}

\maketitle 

\section{Introduction}
\begin{figure}
\includegraphics[width=\columnwidth]{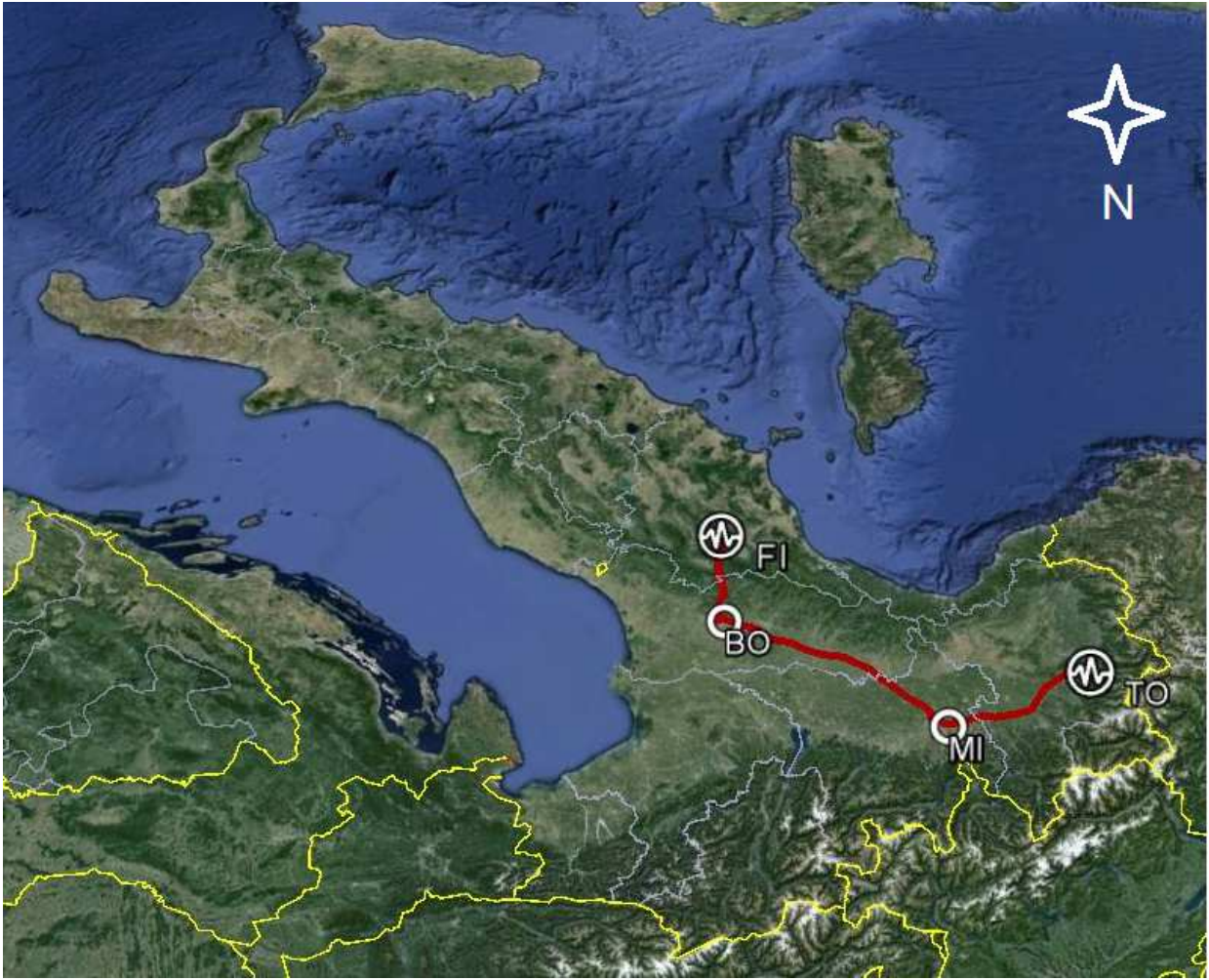}
\caption{\label{fig1} The map of the coherent optical fiber link. View from North to South. TO is Turin, MI Milan, BO Bologna, FI Florence (elaboration based on GoogleEarth).}
\end{figure}

Coherent optical fiber links are the most precise technique to transfer time and frequency signals or to compare remote frequency standards, \cite{predehl,lopez,droste,lopez2,sliwczynski,wang,marra,fujieda,ebenhag,hong,vojtech} as they  improve by more than four orders of magnitude the resolution of current satellite techniques \cite{bauch}. Thus, they are a key technology both for science and metrology, allowing for different outstanding applications.\\
 Presently, optical links are the only viable method for optical clocks comparison. This is a prerequisite for an effective secondary representation of the second in the International System of units \cite{bipm}, as recommended by the Conf\'{e}rence G\'{e}n\'{e}rale des Poids et Mesures, and for the possible redefinition of the second itself.\\
 Even Cs fountains, presently the best realizations of the SI second, could benefit from optical fiber links as an alternative to satellite techniques \cite{bauch}. In fact, optical links would reduce by a factor 20 the measuring time needed to compare primary standards at their accuracy level. \\
Moreover, optical links pave the way for a network of accurate clocks. For instance, in Europe there are about thirty atomic clocks between optical and fountain frequency standards. This network would establish a unique facility for testing fundamental physics, relativistic geodesy and for improving global navigation satellite systems \cite{chou,muller,cerretto}.\\ 
In addition, optical links could improve the synchronization in Very Long Baseline Interferometry (VLBI) antennas and particle accelerators \cite{he,cliche,kim}.\\
Most of these applications require fiber hauls of thousands of kilometers, and this is a challenge for performances, installation costs, and link characterization.\\
This work presents the optical fiber link of \SI{642}{km} implemented in Italy, named LIFT (Italian Link for Time and Frequency) \cite{levi}. LIFT connects the Italian metrological institute (INRIM) to several Italian scientific poles which would largely benefit from improved frequency references. In particular, as shown in the map in Fig. \ref{fig1}, the fiber link provides a reference signal to the Institute for Photonics and Nanotechnologies in Milan, to the Institute for Radio-Astronomy in Medicina (Bologna) and to the University of Florence-European Nonlinear Spectroscopy Laboratory (UNIFI-LENS) in Florence for precision measurements in atomic and molecular physics. LIFT will also connect Italy to the forthcoming European fiber network. \\
To characterize the ultimate performance of the link in terms of residual  phase noise and frequency uncertainty at the far fiber end, we need to compare the frequency of the delivered and of the original signal. We do it by looping the link so that the signal goes from INRIM to UNIFI-LENS and then comes back to INRIM, after travelling \SI{1284}{km} in the optical fiber. In particular, we demonstrate that it is not necessary to use two independent fibers for this characterization. This avoids the need of arming a separate twin link  for the bare characterization purpose: the number of amplifiers is halved, and the reduction of the costs and of the infrastructure complexity benefits the effective implementation of continental long hauls.\\
Different approaches have been proposed in the literature to analyse the link instability and inaccuracy \cite{lopez,droste}. We contribute to this discussion, also addressing some relevant issues in the use of the Allan deviation estimator. In particular, its theoretical evaluation from the observed phase noise spectrum, and the need of reducing the measurement bandwidth, as an alternative to the use of the modified Allan deviation \cite{droste}. This is especially relevant with high phase noise or low control bandwidth, typical in  long haul optical links.\\
\section{Experimental setup}
\begin{figure*}
\resizebox{0.75\textwidth}{!}{%
\includegraphics{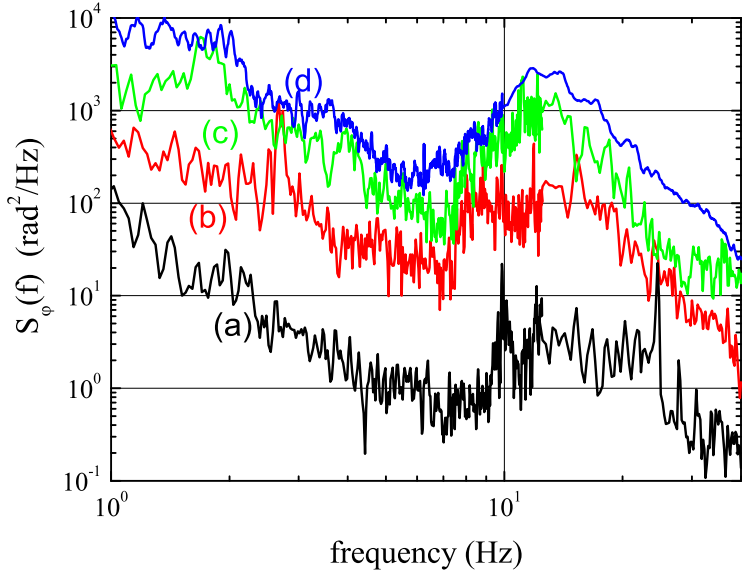}
}
\caption{ The measured phase noise of the hauls between INRIM and some relevant places along the backbone: (a, black) Turin (city), \SI{25}{km}; (b, red) Novara, \SI{169}{km}; (c, green) Piacenza, \SI{347}{km}; (d, blue) Florence, \SI{642}{km}. \label{figN}}
\end{figure*}

The infrastructure is based on a dedicated fiber \SI{642}{km} long, with \SI{171}{dB} losses compensated by 9 bidirectional Erbium Doped Fiber Amplifiers (bEDFA). For a \SI{18}{dB} gain and  on a bandwidth of \SI{0.1}{nm}, these devices exhibit an Amplified Spontaneous Emission (ASE) of \SI{-35}{dBm} at \SI{0.5}{nm} from the coherent carrier at \SI{1542.14}{nm}, of \SI{-38}{dBm} at \SI{1561.4}{nm} and of \SI{-28}{dBm} at \SI{1529.6}{nm}. This large ASE  is filtered on a bandwidth of \SI{0.8}{nm} by 8 telecom optical filters to prevent the amplifiers gain saturation and the onset of auto-oscillations \cite{delisle}. 
Table~\ref{tab1} shows the amplifiers shelters location and the optical loss/gain for each fiber span, and  Fig. \ref{figN} shows the measured phase noise on the haul between INRIM and some intermediate locations along the backbone.  \\
\begin{table}
\caption{The amplifiers placement and the losses/gains for each span.\label{tab1}}
\begin{tabular}{llllll}
\hline\noalign{\smallskip}
 & From: & \textrm{To:} & \textrm{Length} &\textrm{Loss} & \textrm{Gain} \\
&  &  &  /km & /dB & /dB \\
\noalign{\smallskip}\hline\noalign{\smallskip}
1&INRIM  & Turin (city) & 25 & -9 & \\
2&Turin (city) & Santhi\`{a} & 67&  -18 & 19\\
3&Santhi\`{a} & Novara  & 77& -18 & 16 \\
4&Novara& Lainate & 50 &-15  & 13\\
5&Lainate & Milan & 60 & -18 &20\\
6&Milan & Piacenza & 67 &-16 & 17 \\
 7&Piacenza & Reggio Emilia & 94 & -23 & 19 \\
 8&Reggio Emilia & Bologna & 74 & -19 & 16\\
 9&Bologna & Rioveggio& 38  & -10 & 17 \\
10&Rioveggio & Florence & 72 & -18  & 19\\
11&Florence & UNIFI-LENS & 18 & -7 & \\
\noalign{\smallskip}\hline
 & \textbf{Total} & & \textbf{642}  & \textbf{-171} & \textbf{156} \\
 \noalign{\smallskip}\hline
\end{tabular}
\end{table}
\begin{figure*}
\resizebox{0.75\textwidth}{!}{%
\includegraphics{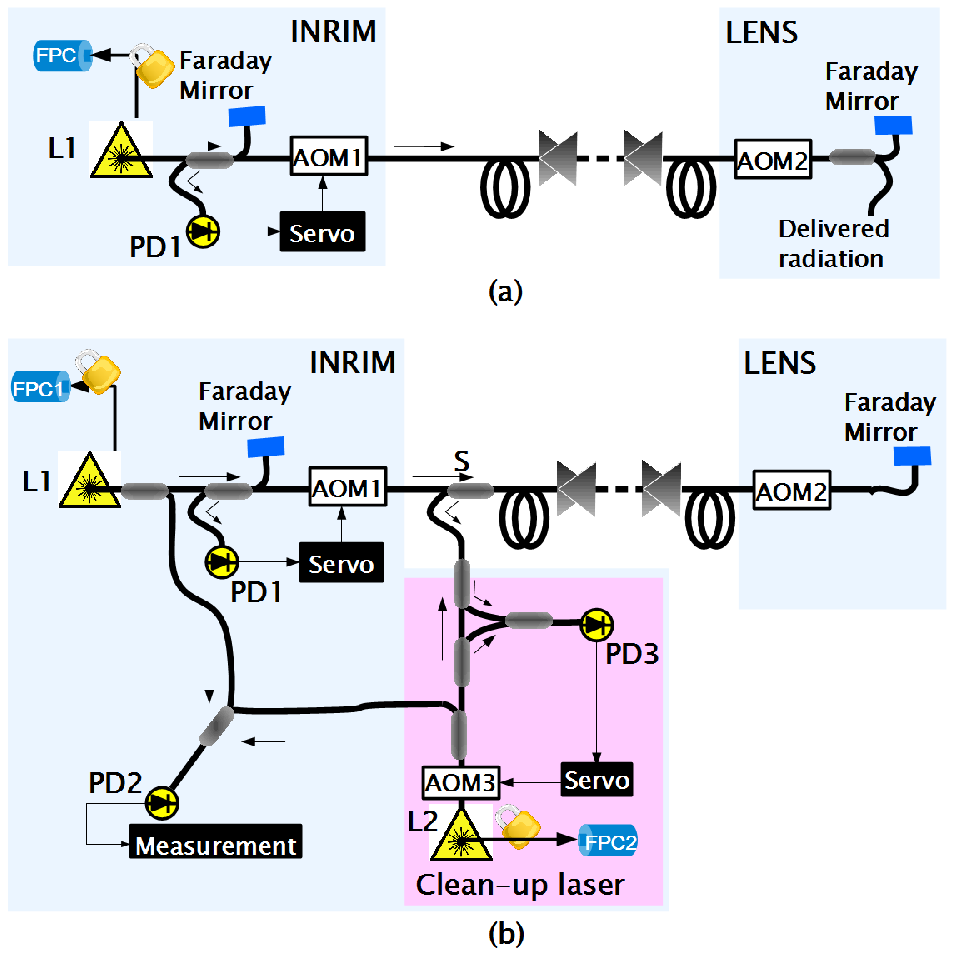}
}
\caption{ (a) The setup of the link Turin-Florence. The radiation from laser L1 is frequency-locked to a Fabry-P\'{e}rot cavity (FPC1) and injected in the link. AOM1 and AOM2 are the local and remote acousto-optic modulators, PD1 is the photodiode used to detect the fiber noise. Triangles represent bEDFA. (b) The setup of the doubled link. The laser coming back from Florence is extracted and regenerated through the clean-up laser L2. L2 is pre-stabilized on a Fabry-Perot cavity (FPC2) then phase-locked to the round-trip signal; AOM3 is the actuator. A part of the radiation from L2 is injected in the link towards Florence for the second round-trip; the remainder is phase-compared to L1 on PD2. \label{fig2}}
\end{figure*}
The delivered optical radiation is provided by a fiber laser at \SI{1542.14}{nm}, frequency-stabilized to an ultrastable Fabry-Perot cavity that reduces the laser linewidth to \SI{30}{Hz} \cite{clivati}. To avoid the stability deterioration of the delivered signal due to temperature variations and mechanical stresses of the fiber, the link is phase-stabilized through the Doppler noise cancellation technique \cite{ma,williams}. The setup is shown in Fig. \ref{fig2}(a). The ultrastable laser L1 is the local oscillator, sent to UNIFI-LENS; few milliwatts of optical power are coupled into the fiber. In the remote laboratory a part of the signal is extracted and represents the delivered ultrastable signal, the remainder is reflected back to INRIM by a Faraday mirror, which compensates for the fiber birefringence. The acousto-optic fixed frequency shifter AOM2, at about \SI{40}{MHz}, is used to distinguish the reflected signal from undesired backreflections along the fiber, that are not frequency-shifted. At INRIM, the round-trip signal is phase-compared  to the local laser on photodiode PD1. This beatnote contains the information about the noise added by the fiber on the round-trip. It is used to stabilize the link with a phase-locked loop (PLL) acting on the acousto-optic modulator AOM1, operating at  about \SI{40}{MHz}.\\
Usually, the characterization of optical links is pursued by arming a second fiber in the same bundle with an independent set of amplifiers \cite{predehl,lopez}. In this work, we demonstrate a characterization technique that uses a single fiber and avoids the doubling of the amplifiers. The setup is shown in Fig. \ref{fig2}(b). The L1 radiation travels to UNIFI-LENS and is reflected back to INRIM, where it is extracted through the splitter S. A second independent ultrastable laser L2 works as a clean-up oscillator for the incoming signal and is phase-locked to it on a bandwidth of about \SI{50}{kHz} by acting on the acousto-optic modulator AOM3, operating at \SI{79}{MHz}. The clean-up radiation travels the doubled link in the backward direction, to provide a double-round-trip signal, needed for the \SI{1284}{km} link stabilization. The clean-up laser is required due to the relevant wideband noise added by the optical amplifiers and enables to achieve a signal to noise ratio of $\sim$\SI{24}{dB} on a bandwidth of \SI{100}{kHz} on the beatnote detected on PD1, which is enough for the doubled link stabilization. The clean-up radiation is up-shifted in frequency by \SI{368}{MHz}  with respect to  the incoming one; the large frequency separation between the signal in each of the 4 passes allows a good rejection of the single Rayleigh scattering, which has a bandwidth of few kilohertz, and avoids any crosstalk between the signals. The link performances are analysed by directly comparing L2 to L1 on the photodiode PD2.
The PD2 beatnote is tracked by a voltage-controlled oscillator and its phase noise spectrum is shown in Fig.~\ref{fig3}, both in the free running (black line) and in the Doppler compensated link configuration (red line). Because the doubled link delay is $\tau=nL/c=$ \SI{6.4}{ms}, with $n=1.468$ the refractive index of the fiber, $c$ the speed of light in vacuum and $L=$\SI{1284}{km}, the loop bandwidth is about \SI{39}{Hz}. The noise of the compensated link achieves the fundamental limit (blue line) set by the round-trip delay \cite{williams}  and demonstrates that uncontrolled effects are not observed at this level. We calculated the round-trip delay limit following the approach described in \cite{williams}, where it is demonstrated that  $S_\varphi(f)= a(2\pi f \tau)^2 S_\text{fiber}(f)$, being $S_\text{fiber}(f)$ the noise of the fiber haul. 
 We calculated $a= 1/4$ rather than $a=1/3$, since in our case a single fiber is used instead of two independent ones; a detailed derivation is given in the appendix.\\
\begin{figure}
\includegraphics[width=\columnwidth]{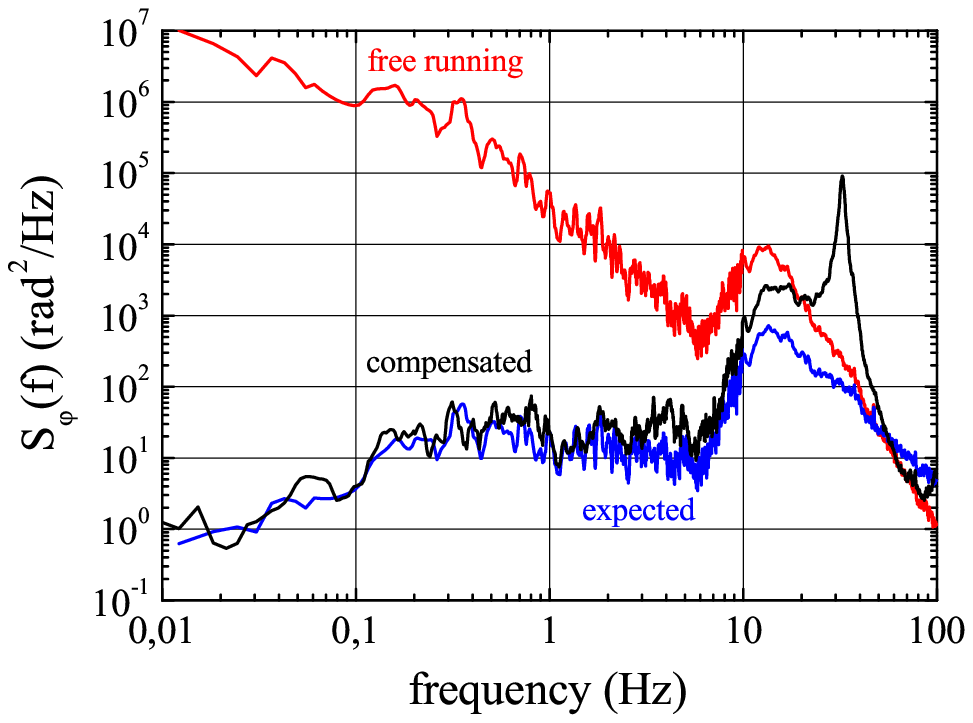}
\caption{\label{fig3}Phase noise spectrum of the free running (black line, a) and compensated (red line, b) \SI{1284}{km} optical fiber link, and the residual phase noise expected from theory (blue line, c). }
\end{figure}
In order to optimize the link performances the gain of each amplifier is tuned to notably reduce the double backreflections along the fiber, which would otherwise dominate the phase noise of the delivered signal.\\
Another detrimental effect is the loss of phase coherence (cycles slips) \cite{udem} on the PLLs, that affects both the instability and the inaccuracy of the link. The cycles slips rate mainly depends on the signal to noise ratio of the beatnotes. In our setup, the cycles slips are a few per hour, provided a polarization adjustment every few hours. 
The phase of the beatnote on PD2 is measured with a dead-time-free digital phase recorder \cite{kramer} at the sampling rate of \SI{1}{ms}, equivalent to a \SI{500}{Hz} bandwidth filter. The cycles slips are barely visible on the phase data even when low-pass filtered at \SI{1}{Hz} bandwidth, as shown in Fig. \ref{fig4} (red line); however if we filter the data on a bandwidth of \SI{0.05}{Hz} (black line) their detection  is quite easy. \\
Particular care is devoted to the filtering process: instead of applying a simple average  to the raw data, that acts like a first order low pass filter with cut-off frequency at half the reciprocal of the averaging time, we implemented a digital Finite Impulse Response (FIR) filter \cite{mitra}, with an out-of-bandwidth attenuation \SI{>70}{dB} between \SIrange{10}{100}{Hz}. The sharper filtering  allows a factor 2 of improvement in the stability for the same equivalent bandwidth, thanks to a higher attenuation of the large noise bump between \SI{10}{Hz} and \SI{100}{Hz}, that is typical in optical links.\\
\begin{figure*}
\resizebox{0.75\textwidth}{!}{%
\includegraphics{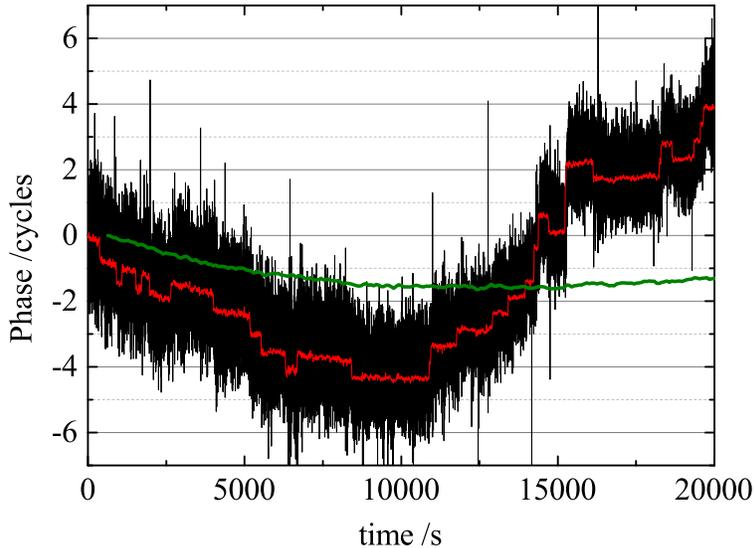}
}
\caption{\label{fig4} The phase of the optical carrier as measured on the photodiode PD2 after travelling the doubled link, filtered on a bandwidth of \SI{1}{Hz} (red line) and \SI{0.05}{Hz} (black line). Data after the removal of the cycles slips are also shown, on a bandwidth of \SI{5}{mHz} (blue line).}
\end{figure*}
We realign the phase by subtracting an integer number of \SI{0.5}{} cycles, i.e. the minimum slip amplitude. Finally, the phase data are further filtered on a \SI{5}{mHz} bandwidth  (smooth blue line in Fig. \ref{fig4}), decimated according to the Nyquist theorem, and differentiated to obtain the instantaneous frequency.\\ 
To evaluate the frequency instability, the Allan deviation estimator is used \cite{allan}. In the literature, its expression is derived from the phase noise spectrum only in the cases where it is modelled by the relation  \cite{rubiola}: $$S_\varphi(f)= \sum_{i=-4}^0 b_\text{i}f^\text{i}.$$ 
An analytical expression is not reported for a phase noise spectrum of the kind  $S_\varphi(f)=b_1 f$, such as the one observed in Fig. \ref{fig3}. This has already been reported in optical fiber links \cite{droste} and depends on the peculiar transfer function of the system, affected by the fiber  delay \cite{williams}. It is important to note that in real systems, a low-pass filter with cut-off frequency $f_\text{h}$ is introduced to prevent noise divergence at high frequency. 
Assuming that the filter is infinitely sharp, we calculate the analytical expression of the Allan variance in presence of the observed type of noise, by performing the integration: 
\begin{equation}
\label{eq1}
\sigma_y^2 (t_\text{a})=\int_0^{f_\text{h}} =h_3f^3 \vert H_\text{A} (f)\vert^2\, df 
\end{equation}
where $t_\text{a}$ is the averaging time and  $\vert H_\text{A} (f) \vert^2 = 2 \frac{\sin^4({\pi f t_\text{a}})}{(\pi f t_\text{a})^2}$ is the squared modulus of the Allan variance transfer function \cite{rubiola}.
For $t_\text{a} \gg 1/(2 \pi f_\text{h})$ the integration leads to:
\begin{equation}
\label{eq1b}
\sigma_y^2(t_\text{a})= \frac{3 h_3f_\text{h}^2}{8 \pi^2} \frac{1}{ t_\text{a}^2},
\end{equation}
During the link operation, we measured an Allan deviation of $\sigma_y(t_\text{a})=8 \times 10^{-13} /t_\text{a}$ on a \SI{100}{Hz} measurement bandwidth and of  $\sigma_y(t_\text{a})=8 \times 10^{-14} /t_\text{a}$ on a \SI{10}{Hz} measurement bandwidth. This is in agreement with the prediction of eq. (\ref{eq1b}) for the phase noise power spectrum observed in  Fig. \ref{fig3}.\\
Figure \ref{fig5} shows the Allan deviation of the free running and compensated optical link on a \SI{1}{Hz} measurement bandwidth, together with the stability of the compensated link on a \SI{5}{mHz} bandwidth. A shorter data sample with less cycle slips was chosen to evaluate the stability at \SI{1}{Hz} measurement bandwidth, since the cycle-slips  could not be efficiently realigned with this bandwidth. The link achieves a short term instability of \SI{1e-14}{} at \SI{1}{s} in a \SI{1}{Hz} bandwidth and an ultimate frequency resolution of \SI{3e-19}{} in a \SI{5}{mHz} bandwidth.\\
We evaluate the accuracy of the link at \SI{5e-19}{} because a non-repeatable frequency offset at this level exceeds the obtained frequency resolution. In spite of this, the present frequency dissemination accuracy is beyond the most challenging requirements of any application and exceeds the world best oscillators \cite{bloom,katori}. \\
\begin{figure}[b]
\includegraphics[width=\columnwidth]{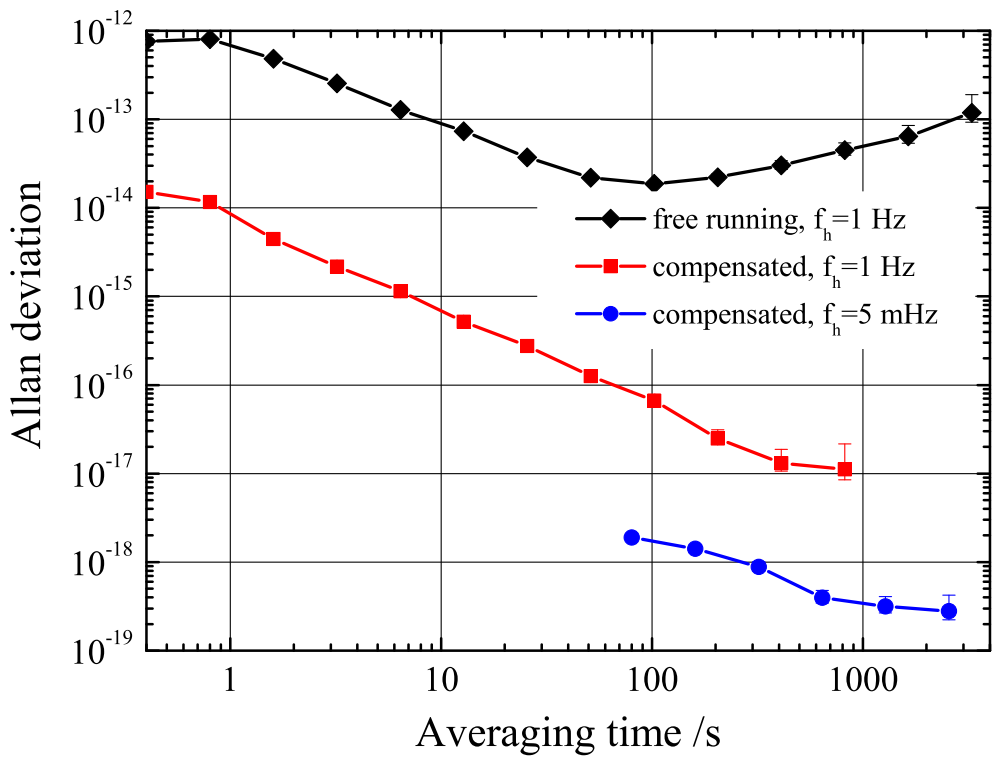}
\caption{\label{fig5}The frequency instability, expressed as the Allan deviation, of the 1284 km optical link in free running (a, black) and compensated (b, red) operation with 1 Hz measurement bandwidth, and the Allan deviation of the compensated link with 5 mHz measurement bandwidth (c, blue). Data have been decimated according to the Nyquist theorem.}
\end{figure}
\section{Conclusions}
This work describes the realization and characterization of a coherent optical fiber link for frequency transfer over a \SI{642}{km} haul. The characterization is pursued by doubling the link length to  \SI{1284}{km}, so that both ends are in the same laboratory.  
The loop has been established using a single fiber; the expected  delay-unsuppressed noise is only slightly affected with respect to the value calculated for independent fibers \cite{williams}. We avoided the arming of a second link; this could be helpful, especially in view of long fiber hauls, as the number of optical amplifiers is considerably reduced. \\
In the data analysis, we addressed the use of the Allan deviation associated to narrow-bandwidth filtering, which allows a resolution equivalent to that obtained using other statistical estimators \cite{droste}.\\
On the doubled link, we obtain a short term Allan deviation of \SI{1e-14}{} at \SI{1}{s} on the bandwidth of \SI{1}{Hz} and an ultimate accuracy on the frequency transfer of \SI{5e-19}{} at \SI{1000}{s} integration time. This is also an upper limit for the real dissemination over the \SI{642}{km} link. \\
The present infrastructure is now being used for the delivery of a high-accuracy frequency reference at LENS, and will benefit the atomic physics experiments which are performed in this laboratory. The haul will soon be upgraded to perform the frequency dissemination  to the Institute of Photonics and Nanotechnologies in Milan and to the Medicina Radio-Telescopes, near Bologna, for the VLBI antennas synchronization. Extraction topologies such as those reported in \cite{bercy,luiten} are being considered  as well as alternative techniques for the remote frequency comparisons \cite{calosso}, and Raman optical amplification \cite{clivatiR} is being investigated as an alternative to the present architecture based on bEDFA. \\

\section{Acknowledgements}
We thank Gesine Grosche and Paul-Eric Pottie for technical help, Giorgio Santarelli for useful discussions, and the GARR Consortium for technical help with the fibers.
This work was supported by: the Italian Ministry of Research MIUR under the Progetti Premiali programme and the PRIN09-2009ZJJBLX project; the European Metrology Research Programme (EMRP) under SIB-02 NEAT-FT. The EMRP is jointly funded by the EMRP participating countries within EURAMET and the European Union.
\section{Appendix A}
This appendix demonstrates the formula for the delay-unsuppressed noise on a link with both ends in the same laboratory, and where the same fiber is used to loop the link, instead of two independent fibers.
Let us write the expression for the round-trip optical phase $\varphi_\text{rt}(t)$ and for the forward signal optical phase $\varphi_\text{fw}(t)$, as a function of the fiber noise $\varphi(z,t)$ at time $t$ and position $z$ along the fiber and as a function of  the phase correction needed for the link stabilization $\varphi_\text{c}(t)$. \\
\begin{equation}
\label{eqAp1}
\begin{split}
\varphi_\text{rt}(t)&=\varphi_\text{c}(t- 2\tau) +\varphi_\text{c}(t)\\
&+ \int_0^{L/2} \Big ( \varphi(z,t-2 \tau+ n\frac{z}{c})+\varphi(z,t- \tau- n\frac{z}{c}) \\
&+\varphi(z,t- \tau+ \frac{z}{c})+\varphi(z,t- n \frac{z}{c}) \Big )  \, dz \\ 
\varphi_\text{fw}(t)&= \varphi_\text{c}(t- \tau)\\
&+\int_0^{L/2} \Big (\varphi(z,t- \tau+ n \frac{z}{c})+\varphi(z,t- n\frac{z}{c}) \Big ) \, dz \\ 
\end{split}
\end{equation}
where $L$ is the loop length, in our case $L=$\SI{1284}{km},  and $\tau=nL/c$ is the link delay. For the sake of clarity, we integrate the length only between $0$ and $L/2$, as the two fiber halves are indeed the same fiber travelled in opposite directions.\\
Let us now assume that the fiber perturbations evolve linearly with time; this is justified for perturbations which act on timescales much longer than $\tau$, as in the case of interest in this context. Within this approximation, eq. \ref{eqAp1} is simplified into
\begin{equation}
\label{eqAp2}
\begin{split}
\varphi_\text{rt}(t)& = 2\varphi_\text{c}(t- \tau) + \int_0^{L/2} 4 \varphi(z,t- \tau) \, dz\\
\varphi_\text{fw}(t)& = \varphi_\text{c}(t- \tau) +\int_0^{L/2} 2 \varphi(z,t- \frac{\tau}{2}) \, dz\\
\end{split}
\end{equation}
Now, considering that in the closed feedback loop configuration $\varphi_\text{rt}(t,z)=0$, eq. \ref{eqAp2} is rewritten as 
\begin{equation}
\label{eqAp3}
\begin{split}
\varphi_\text{fw}(t)&=2\int_0^{L/2}\Big ( \varphi(z, t-\frac{\tau}{2})-\varphi(z,t-\tau) \Big) \, dz \\
&=2\int_0^{L/2} \frac{\tau}{2} \frac{d}{dt} \varphi(z,t-\tau) \, dz\\
\end{split}
\end{equation} 
In the last equation, the evolution of the fiber noise is expressed as a function of its time derivative. For the fundamental theorem of the signal analysis,  \cite{papoulis} the noise power spectrum of the output of a linear and time-invariant system can be written in terms of the  noise of the input; in our case, this theorem can be applied to each fiber segment separately, i.e.:
\begin{equation}
S_\varphi(z,f) = \vert H(z,f) \vert ^2 S_\text{fiber}(z, f),
\end{equation}
where $S_\varphi(z,f)$ is the contribution of a fiber segment with length $dz$ to the compensated forward signal  phase noise, $H(z,f)=\mathcal{F}( \tau \frac{d}{dt})=2\pi i f \tau$ and $S_\text{fiber}(z, f)$ is the phase noise power spectrum of each fiber segment. Assuming that the contributions of each fiber segment are independent, we can perform the integration and end up with the stated result that 
\begin{equation}
S_\varphi(f) =\frac{1}{4}(2 \pi f \tau)^2 S_\text{fiber}(f),
\end{equation}
where $S_\text{fiber}(f)$ is the phase nosie of the \SI{1284}{km}-long link, and it has been used the relation $$S_\text{fiber}(f)=4 \int_0^{L/2} \vert \varphi(z, t) \vert^2 \,dz.$$

\end{document}